\documentclass[12pt, authoryear]{article}

\usepackage[latin1]{inputenc}
\usepackage{graphicx, amsmath, amsthm, amsfonts, amssymb, subfigure, amstext}
\usepackage{wrapfig, amsopn, latexsym,  epsfig, multicol, multirow,verbatim}
\usepackage{comment,  caption,  bm}
\usepackage[margin=1in]{geometry}

\title{Bayesian  inference for  asymptomatic COVID-19 infection rates}
\author{Dexter Cahoy$^{\S}$, Joseph Sedransk$^{\dagger}$} 

\date{}

\usepackage{natbib} 
\usepackage[plainpages=false,pdfpagelabels,hypertexnames,linktocpage, colorlinks, citecolor=red,linkcolor=blue]{hyperref}

\begin{document}

\setlength{\parindent}{0pt}
\setlength{\parskip}{7pt}

\maketitle
 
\renewcommand{\thefootnote}{\fnsymbol{footnote}}
\footnote[4]{Department  of Mathematics and Statistics,  University of Houston-Downtown, USA, email: {\tt cahoyd@uhd.edu}}
\footnote[3]{Joint Program in Survey Methodology, University of Maryland, College Park, MD 20742, email: \texttt{jxs123@case.edu}}

\begin{abstract}
\indent

\vspace{0.1in}To strengthen inferences meta-analyses are commonly used to summarize information from a set of independent studies. In some cases, though, the data may not satisfy the assumptions underlying the meta-analysis. Using three Bayesian methods that have a more general structure than the common meta-analytic ones, we can show the extent and nature of the pooling that is justified statistically.  In this paper, we reanalyze data from several reviews whose objective is to make inference about the COVID-19 asymptomatic infection rate.  When it is unlikely that all of the true effect sizes come from a single source researchers should be  cautious about pooling the data from all of the studies. Our findings and methodology are applicable to other COVID-19 outcome variables, and more generally.

\noindent \textbf{Keywords}:  Dirichlet process mixture, exchangeable random variables,   meta-analysis, pooling results, reversible jump Markov Chain Monte Carlo,  SARS-CoV-2

\end{abstract}

\section{Introduction} \label{sec1}

Meta-analyses are commonly used to summarize information from a set of independent experiments, observational studies or sample surveys. Doing this may strengthen inferences when there are deficiencies in the individual studies such as small sample sizes. Methodology for combining findings from repeated research studies has a long history and, in particular, meta-analyses have become very popular over the past thirty years. From an online search for `books meta-analysis' we found forty-nine books. Thus, it was natural that early in 2020 several meta-analyses were conducted (and subsequently published) about infection rates from the novel coronavirus.  Looking at several early review papers
we were concerned whether the meta-analyses were carried out in an appropriate manner.
Even after careful evaluation to include only studies thought to be comparable, there may be subsets of the collection of studies where the true (subset) effects are very different.  If this is so, pooling the data from all of the studies may result in misleading conclusions.  \cite{bhh10} add: ``If the variation is substantial, then we might want to shift our focus ....
Rather it should be on the fact that the ... effect differs from study to study. Hopefully, it would be possible to identify reasons ...
that might explain the dispersion."

In this paper we consider three Bayesian methods that have a more general structure. One can use these methods to check the validity of the more standard approaches by investigating  whether the set of true effect sizes come from a common source. If the assumptions underlying the standard approaches are not met, our proposed methodology will lead to more appropriate inferences.

From five review papers we selected for further analysis several studies that have different features. In each of these cases the objective is to make inference about the asymptomatic infection rate.

Please note that we are not evaluating specific meta-analytic methods. Our concern is about appropriate aggregation of possibly disparate data.

\cite{bge20} explain the importance of a review: ``Accurate estimates of the proportions of true asymptomatic and presymptomatic infections are needed urgently because their contribution to overall SARS-CoV-2 transmission at the population level will determine the appropriate balance of control measures. If the predominant route of transmission is from people who have symptoms, then strategies should focus on testing, followed by isolation of infected individuals and quarantine of their contacts. If, however, most transmission is from people without symptoms, social distancing measures that reduce contact with people who might be infectious should be prioritized, enhanced by active case-finding through testing of asymptomatic people.''  Referring to a narrative review report  \citep{ot20} that presents a range (over studies) of 6\% to 96\% for the proportion of individuals positive for SARS-CoV-2 but asymptomatic, the authors point out the need for a careful review.

Standard meta-analyses typically assume that the true effect sizes, $\mu = (\mu_1,...,\mu_L)^t,$ come from a common source.
Even after including only those studies thought to be comparable, $\mu$ may be composed of distinct
subsets, each with a different underlying distribution.
This seems likely for some of the reviews, e.g., the seventy-nine rates in  \cite{bge20} that range from 0.01 to 0.92. To make appropriate inferences
the three Bayesian methods have a more general structure than that
assumed in a standard meta-analysis. The principal
method, termed \emph{uncertain pooling}, is flexible in that it can identify distinct subsets of $\mu$: e.g., for $r$ subsets there
would be $r$ true effect sizes, $\nu_1,\ldots, \nu_r$. Then, pooling the data from all of the studies
may lead to misleading inferences. This methodology will also indicate when true effect sizes have a common source, thus leading to an appropriate inference. The more general structure should ensure greater concordance of the data with our model than
with a more restricted model.  A better fitting model should lead to better inference. Specifically, only similar studies will be combined.
It is not surprising that there is strong statistical evidence that in three of the four data sets that we analyze (Section 4)
the true effect sizes do not come from a single source. Then the analyst should be  cautious about combining the data from all of the studies.

For a general discussion of Bayesian methods for meta-analysis see \cite{scw21}.  \cite{bhh10} is a basic treatment of fixed and random effects models for a meta-analysis while  \cite{rhl18} re-evaluate fixed effect(s) meta-analysis.

Section 2 has brief descriptions of the data sets that we analyze, together with background information. The
methodology is introduced in Section 3 while the results are summarized in Section 4.
Section 5 has a brief summary
and an extension that accommodates study level covariates with notes about the availability of covariates in the reviews we investigate.

\section{Study Descriptions}

This section has brief descriptions of the meta-analyses that we have analyzed together with some background information.
The definitions of \emph{asymptomatic infection rate} and the conditions required
for including individual studies in the meta-analysis differ considerably, and are too detailed to present all of them in this paper.

The first paper that we considered was  by  \cite{he20}.  Using data from six studies they obtain estimates of the   asymptomatic infection rate, noting that these measures differ considerably over the six studies, and explaining that this may be due to ``differences in data collection, sample size, and the conditions.''  Since the information from one of the six studies is inconsistent with that in the other five studies we include only the latter in our analyses.  As seen in Table 1 the sample proportions range
from 0.22 to 0.78 with little or no clustering. 

The first meta-analysis concerning only the asymptomatic coronavirus disease rate is \cite{heb20}.  In their Section 1 the authors note the importance of studying the asymptomatic rate and that this rate is not ``well characterized.''   They conduct meta-analyses for all 41 studies and five subsets.  The rates are markedly heterogeneous (proportions from 0.02 to 0.75 and total numbers of cases from 4 to 44,627), suggesting concern about the aggregation. Our analysis is for the children subgroup, with eleven studies. As seen in Table 2, the sample proportions range from 0.11 to 0.57 while the SEs range from 0.01 to 0.16. Unlike the first
meta-analysis, there is some apparent clustering.

The third data set is a subset of six of the eleven studies in  \cite{heb20}. These six studies were chosen to illustrate properties
when there is considerable separation. From Table 3, there are two apparent clusters with cluster proportions of about 0.13 and 0.55; overall,
the SEs range from 0.01 to 0.16.

\cite{bge20} has several important features, i.e., the authors consider rates associated with both asymptomatic and presymptomatic cases and only include studies that document follow-up and symptom status at the beginning and end of follow-up or modeling. Their meta-analyses are based on seventy-nine studies, summarized for the entire set and seven subsets.  The seventy-nine rates are markedly heterogeneous (proportions from 0.01 to 0.92 and total numbers of cases from 2 to 1012).  Our analysis is for the screening subgroup with seven studies, noted by \cite{bge20} as being of special interest. Here, the proportions range from 0.17 to 0.50 while the SEs range from 0.04 to 0.35.

For additional background we give the criteria that  \cite{heb20} used to select the studies that they used in their meta-analyses.
Then we summarize features of the eleven children study, from \cite{heb20}, that we have analyzed.   \cite{heb20} searched two databases, PubMed and Embase, following the Preferred Reporting Items for Systematic Reviews and
Meta-Analyses (PRISMA) guideline. They included the following items:
``COVID-19" and analogous phrases and ``Asymptomatic." They included ``articles reporting a specific number of asymptomatic
infection cases in confirmed COVID-19 patients, information describing the epidemiological and clinical features of COVID-19." There is
no evidence of a risk-of-bias assessment or consideration of a sufficient follow-up period.  \cite{bcb20}  identify
these as characteristics essential when making decisions about which studies to include in a meta-analysis. Since  \cite{heb20}
is the first meta-analysis to make inferences about the asymptomatic infection rate, one may conjecture that they were motivated to
publish their results quickly.

Of the eleven studies of children five papers are published in Chinese, so there are only summaries in English. Only three of the eleven
papers give the age distribution of the children, although most give the mean age and some give the range. Seven of the papers give
the sex distribution. In five studies all cases were associated with a single hospital while four studies summarized the results from
many hospitals. There was no information for two studies. Several papers noted that most of the patients had a history of close contact
with adults with COVID-19.

\section{Methodology}%

A common assumption in situations where combining data is plausible is: For $i=1,\ldots,L, j =1,\ldots, n_i,$ the $Y_{ij}$ are independent
\begin{equation}\label{sample mean}
\overline{Y}_i \sim N(\mu_i, \sigma_i^2/n_i)
\end{equation}
where $\overline{Y}_i = \sum_{j=1}^{n_i} Y_{ij}/n_i$, the $\sigma_i^2$ are known, $L$ is the number of studies and  $n_i$ is the number of replicates.
Note that all of the analyses of COVID-19 data that we consider make assumptions like (\ref{sample mean}).

\subsection{Uncertain pooling}

The first method, uncertain pooling, is based on    \cite{mas92}  and  \cite{eas01}. Since this method may be unfamiliar, we describe
it in some detail. They showed
that a prior for $\mu=(\mu_1, \mu_2, \ldots, \mu_L)^t$  can be selected to reflect the beliefs that there are subsets of $\mu$ such that the $\mu_i$ in each subset are ``similar'', and that  there is uncertainty about the composition of such subsets of $\mu.$   Let $G$ be the total number of partitions of the set $\mathcal{L} =\{1,\ldots, L \}.$  Denote a particular partition by $g= 1, \ldots, G$, let $d(g)$ denote the number of subsets of $\mathcal{L}$ in the $g$th partition $(1 \leq  d(g) \leq L)$, and let $S_k(g)$ denote the set of study labels in subset $k$ for $k= 1, ... , d(g).$ For our analyses $L = 5, 6, 7$ and $11$ with $G = 37, G = 203, G = 877$ and $G = 678,570,$ respectively.
For other values of $L$ the total number of partitions of an $L$-element set is given by the Bell number, $B_L.$
Recent work \citep[e.g.,][]{da17} proposed using prior information to place a prior on the set of partitions $g = 1,...,G.$ This will increase
the complexity of the computations but avoid the need to consider the $G$ partitions explicitly.

To specify a prior for $\mu,$  first condition on $g.$    \cite{mas92}  and    \cite{eas01}  assume that there is independence between  subsets, and within $S_k(g)$ the $\mu_i$  are independent with
\begin{equation}\label{first stage prior}
\mu_i \; | \; \nu_k(g)  \sim N (\nu_k(g), \delta_k^2(g))   \qquad i \in S_k(g).
\end{equation}
Also, the $\nu_k(g)$ are mutually independent with
$$\nu_k(g) \; | \; \theta_k(g)  \sim N (\theta_k(g), \gamma_k^2(g)).$$
Conditioning on the variances above (but suppressing them in our notation), and letting $\gamma_k^2(g) \to \infty$ leads to the following expected results.
Defining $y = {(\bar{Y}_{1},\ldots,\bar{Y}_{L})}^{t}$ and letting $\Delta = \{\delta^2_k(g): k = 1,\ldots,d(g); g =
1,...,G\}$
\begin{equation}
\label{posterior expectation}
E(\mu_i | y,g, \Delta ) = \{\lambda_i (g) \} \hat{\mu}_i + \{ 1-\lambda_i (g) \}  \hat{\mu}_k(g)  \qquad  i \in S_k(g)
\end{equation}
and
\begin{equation}
\label{posterior covariance}
{\text{cov}} (\mu_i, \mu_t | y, g, \Delta) = \left.
\begin{cases}
  \delta_k^2 (g) \lbrace 1-\lambda_i (g) \rbrace   + \frac{ \lbrace 1-\lambda_i (g)  \rbrace^2  \delta_k^2 (g) }{\sum_{i \in S_k(g)} \lambda_i (g)},   &   i=t; i,t \in S_k(g) \\
 \frac{ \lbrace 1-\lambda_i (g) \rbrace \lbrace 1-\lambda_t (g) \rbrace \delta_k^2 (g) }{\sum_{i \in S_k(g)} \lambda_i (g)},    &   i\neq t; i,t \in S_k(g) \\
 0, &      i  \in S_{k_1} (g), t  \in S_{k_2} (g), k_1 \neq k_2, \\
 \end{cases}
\right.
\end{equation}
where
\begin{equation}\label{lambda}
\lambda_i(g) =    \frac{\delta^2_k(g) }{  \delta^2_k(g)  + \sigma_i^2/n_i },  \;  \hat{\mu}_k (g) = \frac{\sum_{t \in S_k(g)} \lambda_t(g) \hat{\mu}_t}{ \sum_{t \in S_k(g)} \lambda_t(g)},  \;
  \text{and} \;
\hat{\mu}_i=\overline{Y}_i.
\end{equation}

Inference about $\mu$ includes uncertainty about the value of $g,$ i.e.,
\begin{equation}\label{posterior mu}
f(\mu | y) = \int f(\mu | y, g, \Delta)f(g, \Delta | y)dg d\Delta
\end{equation}
where the notation is simplified by using integration rather than summation for $g.$


To evaluate (\ref{posterior mu}) we need $f(g, \Delta | y).$  One must be careful about
specifying how the $\gamma^2_k(g)  \to \infty$ because the models corresponding to the partitions have different numbers of parameters. We use a method described in Section 3 of  \cite{eas01} that postulates little prior, relative
to sample, information about the $\nu_k(g)$ and is invariant to changes in the scale of $Y$. Let $\nu(g) = (\nu_1(g),...,\nu_{d(g)}(g))^t$
and $K(f_1(\nu(g)), f_2(\nu(g)| y))$ be the Kullback-Leibler information about $\nu(g)$. With
prior $f(g,\Delta) = f(g)f(\Delta)$ and letting the $\gamma_k^2(g) \to \infty$ subject to $K(f_1(\nu(g)), f_2(\nu(g)| y))$ = constant
\begin{align}\label{posterior g, delta}
f(g, \Delta |y)  \propto & \; f(\Delta) f(g)\exp \left\lbrace \frac{-d(g)}{2} \right\rbrace \prod \limits_{k=1}^{d(g)} \prod \limits_{i \in S_k(g)}  \lbrace 1-\lambda_i (g) \rbrace^{1/2} \nonumber \\
&\times \exp \left[ -\frac{1}{2} \sum\limits_{k=1}^{d(g)}  \sum\limits_{i \in S_k(g)} \left\lbrace \frac{\lambda_i(g)}{\delta^2_k(g)}  \right\rbrace \lbrace \hat{\mu}_i - \hat{\mu}_k(g) \rbrace^2  \right],
\end{align}
The factor in the exponent,
$$ Q\{d(g)\} = \sum\limits_{k=1}^{d(g)}  \sum\limits_{i \in S_k(g)} \left\lbrace \frac{\lambda_i(g)}{\delta^2_k(g)}  \right\rbrace \lbrace \hat{\mu}_i - \hat{\mu}_k(g) \rbrace^2, $$
a consequence of the limit process just described, is the usual within sum of squares from a conventional, weighted, analysis of variance.
Now, $Q\{d(g)\}$ is likely to decrease as $d(g)$ increases, for example for a new partition of $\bigcup_{k=1}^{d(g)} S_k(g)$ obtained by creating subsets
of the existing $\{S_k(g)\}$. Since $f(g, \Delta | y)$  increases as $Q\{d(g)\}$ decreases, it is helpful to have the
second term, ${\text{exp}}\{-d(g)/2\}$, that penalizes partitions with larger values of $d(g)$.

For our analysis we take $\delta^2_k(g) = \delta^2$ and write $\lambda_i(g) = \delta^2/(\delta^2 + \sigma^2_i/n_i).$
Inference for $\mu$ is made using (\ref{posterior mu}) and (\ref{posterior g, delta}) with
\begin{equation} \label{posterior mean}
\mu | y, g, \delta^2 \sim N(E(\mu | y, g, \delta^2), V(\mu | y, g, \delta^2))
\end{equation}
where the conditional posterior moments of $\mu$ are given in (\ref {posterior expectation}) and (\ref{posterior covariance}).


Our analyses will indicate
whether the true effect sizes come from a common source.
If so, then using a standard meta-analysis will provide appropriate inference. If not, several alternatives should be considered, as discussed below.

If a prior evaluation
indicates that one of the studies, $i$, can be regarded as a gold standard we can consider the posterior distribution corresponding to study $i$ to be the object of
inference. Then, using the posterior expected value for illustration,
$$E(\mu_i | y) = E_{\delta^2, g | y}E(\mu_i | y, g, \delta^2)$$
where $E(\mu_i | y, g, \delta^2)$ is defined in (\ref{posterior expectation}). Thus, inference for $\mu_i$ is a function of $\hat{\mu_i}$
together with data from the other $L - 1$ studies as determined by the form of (\ref{posterior expectation}), and, critically, by the likelihood
associated with the set of subsets, $S_k(g)$, containing study $i.$  See   \cite{eas01} for additional details and an application to a notable study of the effect of using aspirin by patients following a myocardial infarction.


Otherwise, one must rely on substantive evaluation to decide whether any distinct subsets identified in the analysis should
be analyzed separately, e.g., by separate standard meta-analyses. If there are no covariates that can distinguish the subsets, then
the distribution of the true effect for study $i$ is a mixture distribution with unknown probabilities associated with
the components.

If there are distinct subsets and a single analysis is presented it is important to include a credible interval for the overall
true effect. Taken together with the presence of
distinct subsets, a very wide interval for the overall true effect would be a strong indication that the single
summary rate is not informative.

\subsection{Dirichlet process mixture}

An alternative to the uncertain pooling method is to use a Dirichlet process mixture (DPM), one of the most popular nonparametric Bayesian methods. This methodology, presented in detail in Sections 2.1 and 2.2 of  \cite{mqjh15} is outlined below, paraphrasing the text in \cite{mqjh15}.  Suppose that there is an observed i.i.d. sample $$y_i | H \stackrel{iid}{\sim} H, \quad i = 1,...,n.$$ For Bayesian inference a prior probability model on $H$ is needed.
Starting with the basic Dirichlet process (DP), let $M > 0$ and $H_0$, the base distribution, be a probability measure defined on the sample space $\mathcal{S}$.
A DP with parameters $(M, H_0)$, is a random probability measure $H$ defined on $\mathcal{S}$ which assigns probability $H(B)$ to every
measurable set $B.$ The DP is an infinite-dimensional analogue of the finite-dimensional Dirichlet prior. In particular, $E(H(B)) = H_0(B)$.
If $M$, the precision parameter, is large $H$ is highly concentrated around $H_0$; as $M \to \infty$ the process is essentially $H_0.$

With DP $H$ is a discrete
measure, so it is typical to extend the DP to DPM, by using a mixture over a simple parametric form such as a $N(\mu, \sigma^2)$ pdf.
Let $\Theta$ be a finite dimensional parameter space. For each $\theta \in \Theta$, let $f_{\theta}$ be a continuous pdf. Given a probability
distribution $H$ defined on $\Theta$, a mixture of $f_{\theta}$ with respect to $H$ has the pdf
$$f_H(y) = \int f_{\theta}(y) dH(\theta).$$ This mixture model can be expressed as an equivalent hierarchical model, especially
relevant for our application, i.e.,
\begin{equation}\label {hierarchy first stage}
y_i | \theta_i \stackrel{iid}{\sim} f_{\theta_i}
\end{equation}
\begin{equation}\label{hierarchy second stage}
\theta_i | H \stackrel {iid}{\sim} H
\end{equation}
with $H \sim DP(M, H_0)$.


For our analyses have used the R function DPmeta from the package \texttt{DPpackage}: see  \cite{jhqmr11} for additional details.

The model in DPmeta, for $\sigma_{i}^2$ fixed, is (\ref{hierarchy first stage}) and (\ref{hierarchy second stage})
with $y_i = \overline{Y}_i, \theta_i = \mu_i, f_{\theta_i}$ the pdf of a $N(\mu_i, \sigma^2_i/n_i)$ random variable,
and $H_0 = N(\eta, \tau^2).$

The (independent) hyperparameters are
$$M | a_0, b_0 \sim \text{Gamma} (a_0, b_0)$$
$$\eta | \eta_b, S_b \sim N(\eta_b, S_b)$$
$$\tau^{-2} | \phi_1, \phi_2 \sim \text{Gamma}({\phi_1}/2, {\phi_2}/2).$$


The uncertain pooling method requires only that one specify a prior distribution for $g$ and $\delta^2.$
By contrast DPmeta requires substantial prior input, i.e., values for $a_0, b_0, \eta_b, S_b, \phi_1$ and $\phi_2.$ We have concerns about the sensitivity of inferences to some of the
choices of distributions in DPmeta and possible over-fitting since there are more quantities to be specified than in the uncertain
pooling method. Moreover, our analyses include data from only five, six, seven and eleven studies. So, we have omitted the specification
$M | a_0, b_0 \sim {\text{Gamma}} (a_0, b_0)$ and made inferences for a selected set of values of $M$ as suggested by  \cite{mde94}.
Also, we omitted the step, $\eta |\eta_b, S_b$, and replaced $\eta$ and $\tau^2$  with their maximum likelihood estimates.

\subsection{Binomial-beta model and reversible jump Markov chain Monte Carlo}

As noted by a reviewer, a limitation of our uncertain pooling method is that the sample standard errors are assumed to be known, as
is typically done. There is contemporary research   \citep{yao21}
that models both the sample mean and the log of the sample standard error. However, they assume a bivariate normal distribution
for these two statistics, a questionable
assumption for our (binomial) case. An alternative that we have investigated assumes a binomial likelihood together with beta and uniform
prior distributions in a hierarchical model. The limitation, here, is that one must use Reversible Jump Markov Chain Monte Carlo (RJMCMC) for the computations,
and implementation is substantially more difficult than the method presented in Section 3.1.

Recall that the methodology used in Section 3.1 is based on  \cite{eas01}  who use a constraint in the limit process to overcome the
problem that the partitions, $g$, have different sizes, i.e., that $d(g)$ varies with $g$. Without this adjustment $p(g|y)$ would not be invariant
to changes in the scale of the outcome variable, $Y.$ RJMCMC addresses the problem of varying $d(g)$ by introducing
additional random variables that enable the matching of parameter space dimensions across the partitions. For a concise outline of the
RJMCMC method see  \cite{gel04}.  The pioneering paper  \citep{gre95} provides the theoretical background for RJMCMC, but also includes,
as an example, an application to the uncertain pooling methodology. We have used this model and RJMCMC to analyze the data in the four
tables in Section 4.

Assume $L$ independent responses, $y_1,\ldots,y_L$, with $y_i \sim  \text{bin}(n_i, \theta_i).$ Within $S_k(g)$
\begin{equation}\label{betabinomial}
\theta_i \stackrel{iid}{\sim} \text{beta}(q\alpha_k, q(1 - \alpha_k)) \quad i \in S_k(g); \quad k = 1,...,d(g).
\end{equation}

The group mean parameters, $\{\alpha_k\}$, are drawn independently from $U$(0,1) while log $q$ $\sim U$(log $a$, log $b).$
Finally, $p(g) \propto d(g)^{-1}/ \#\{ g^{'}: d(g^{'}) = d(g)\}$.

Then the joint distribution of all the variables is $p^* = p(g, \alpha,q,\theta,y)$ with
\begin{eqnarray*}
 p^* &=& p(g)p(\alpha, q |g)p(\theta |g,\alpha, q)p(y|g, \alpha, q, \theta) \\
                                           & =& p(g)p(\alpha | g)p(q)p(\theta |g,\alpha,q)p(y|\theta) \\
                                            &=& p(g)\Big[\prod_{k=1}^{d(g)}1\Big] p(q) \prod_{k=1}^{d(g)}\prod_{i \in S_k(g)}\frac{\theta_i^{q\alpha_k-1}(1- \theta_i)^{q(1-\alpha_k) - 1})}{B(q\alpha_k,q(1 - \alpha_k))}\prod_{i=1}^L {n_i \choose y_i}\theta_i^{y_i}(1- \theta_i)^{n_i - y_i}.
                                         \end{eqnarray*}
where the $1$ in the last expression is the pdf of the (assumed) uniform distribution for $\alpha_k.$

Section 6.2 of  \cite{gre95} gives the full conditionals for $\theta_i, q$ and $\alpha_j.$  The step involving
a possible move from partition $g$ to a new partition $g^{*}$
is much more complicated.  \cite{gre95}  uses a process that jumps between partitions making only the changes of splitting a group (a birth) and
combining two groups (a death). There is an algorithm to select the groups to split and merge. Then births are attempted with probability $b_g$ and deaths with probability $d_g.$ Jumping to a new partition requires a change in the vector $\alpha$ since its length must increase
or decrease by one unit. Several steps are required to develop the associated proposal, finally leading to a complicated acceptance probability.


For the pool-all partition  \cite{ha08} suggest
$$y_i \stackrel{ind}{\sim} \text{bin}(n_i, \theta_i)$$
with $\eta_i = \text{logit} (\theta_i)$ and
$$\eta_i \stackrel{iid}{\sim} N(\xi, \kappa^2).$$
As an alternative it may be feasible to adapt this to the specification in (\ref{betabinomial}),
and implement it using RJMCMC.

\section{Results and Discussion}%

Our inferences are for the asymptomatic rates, i.e., population proportions, and Tables 1 - 4 use this representation. That is,
in Tables 1-4 we present the sample proportions and the
 standard errors (SEs). The
posterior means and intervals from basic uncertain pooling (Section 3.1) and summaries from DPmeta (Section 3.2) and
RJMCMC (Section 3.3) are for the population proportion.
However, for the computations for basic uncertain pooling and DPmeta we have
used two different representations of $\overline{Y}_i$ in (\ref{sample mean}), i.e., $\overline{Y}_i$ is the sample proportion, $\hat{p}_i$,
and $\overline{Y}_i$ is the sample log odds, $\text{log}[\hat{p}_i/(1 - \hat{p}_i)]$. While we only present results  for the latter representation, our findings are similar for both choices.
From (\ref{sample mean}), $(\overline{Y}_{i}, \mu_i, \sigma^2_i/n_i)$ is replaced by
\begin{equation}\label{transformed notation}
(\text{log} [\hat{p}_i/(1 - \hat{p}_i)], \text{log} [p_i/(1 - p_i)], 1/[n_i p_i(1- p_i)]).
\end{equation}
As is typically done in applications such as this, e.g.,  \cite{dal86},  we have replaced $\sigma_i^2$ with an estimate from the sample.  For each of the meta-analyses we give in the Appendix the number of asymptomatic cases and observations for each
of the component studies together with references where these data can be found.

For uncertain pooling inference for $\mu$ is made using (\ref{posterior mu}). To start, evaluate the right side of (\ref{posterior g, delta}) for
$$ \{g: g = 1, . . ., G; \quad  D \; \text{grid points for} \; \delta^2 \;  \},$$
then
standardize by dividing the individual terms in the grid by their sum. This provides an approximation for
$f(g, \delta^2 | y).$
Then select a random sample of size $B$ from the $DG$ normalized values of $f(g, \delta^2 | y).$ For each selection, $(g_{*},\delta^2_{*}),$ 
sample $\mu$
from  $f(\mu | y, g_{*}, \delta^2_{*}).$ For the logit case, transform $\mu$ to $p = (p_1,...,p_L)^t$ at each step.
Starting with a grid for $\delta^2$ and $g$ with a very large range for $\delta^2$, we reduced the $(g, \delta^2)$ space to make
the 2D-grid sampler faster. Specifically, we retained 99.2\% of the probability associated with the extensive grid.
We generated
$B = 30,000$ values of $\mu$ for the eleven study case and $B = 10,000$  for the other cases.
Finally, note that approximations for the marginal posterior distributions, i.e., $f(g | y)$ and $f(\delta^2 | y),$ can be obtained
directly from the grid approximation of $f(g, \delta^2 | y)$.


As noted in Section 1, our study was motivated by a meta-analysis by   \cite{he20} that included
early studies of the COVID-19 asymptomatic infection rate.   \cite{he20}  carried out a standard meta-analysis using a normal-based random effects model with the effect sizes as the outcome random
variable and assuming fixed SEs. Since there is considerable variation in the effect sizes it is
prudent to be cautious and investigate whether the true effect sizes are from a single source.

Our analysis starts with the basic uncertain pooling method (Section 3.1) and DPmeta (Section 3.2). We summarize the results
using the binomial-beta model and RJMCMC (Section 3.3) at the end of this section. For the basic uncertain pooling method we assumed, a priori,
that all 37 partitions have equal probability, i.e., $p(g) = 1/37.$ For the prior on $\delta^2$, independent of $g$, we used two distributions:

\begin{table}[h!t!b!p!]

\centering

\setlength\tabcolsep{2pt}
\renewcommand{\arraystretch}{-1.5}
\begin{tabular}{llc}
a) InvBeta: &  $p(\delta^2) \propto 1/(1 + \delta^2){\surd{\delta^2}},$ &  $0 < \delta^2 < \infty.$ \\
b) InvGamma: & $p(\delta^2) \propto (\delta^2)^{-(\alpha + 1)}$ exp${-(\beta/\delta^2)}, $ & $0 < \delta^2 < \infty.$ \\

\end{tabular}

\end{table}



For a standard random effects model,  \cite{gel06} recommends using a half-Cauchy prior distribution for $\delta,$
$p(\delta) \propto 1/(1 + \delta^2).$
While our situation is very different, i.e., many partitions and weighting depending on $\delta^2$ (not $\delta$), we have adopted this
suggestion in (a) by transforming the half-Cauchy pdf to obtain the Inverse Beta (InvBeta) pdf of $\delta^2.$
Previous research has shown that there are benefits to having the prior distribution for $\delta^2$
concentrated near 0, leading to the choice in (b) with $\alpha = 11.01, \beta = 0.001$. In the following we present results only for (a) as those for (b) are similar.

As described in Section 2, a complete specification of DPmeta requires estimates of many hyperparameters. This seems inappropriate for meta-analyses such as these. So, we have replaced  $\eta$ and $\tau^2$ with their maximum likelihood estimates, thus eliminating the need to specify the other hyperparameters. We have adopted the suggestion of  \cite{mde94} to use a small set of values for $M$, i.e., $\{L^{-1}, L^0, L^1, L^2\}$, typically augmented by a value for $M$ much smaller than $L^{-1}$ and one larger than $L^2.$

We start by discussing the meta-analysis that motivated our investigation, i.e., He, Yi,  and Zhu.\cite{he20}
We analyze the data (sample proportions, numbers of observations) from five of their studies, omitting one  \citep{mi20} whose estimates
are based on extensive modelling. As such,  \cite{bge20} ``considered ... [this] study ... separately, because of the different method of analysis," and  \cite{heb20}  excluded it entirely.   Note that the numbers of asymptomatic cases and observations are included in the Appendix.

In Table 1 we present for each of the five studies the sample effect size (sample proportion, $\hat{p}$), posterior expected value of the true effect size, standard error (SE), and 95\% credible interval for the true
effect size. Here, SE = $\sqrt{\hat{p}\hat{q}/n}$ where $n$ is the total number of observations. From DPmeta, there are the
posterior expected values of the true effect size corresponding to $M = 1/5$ and $5$  (a good representation of the
six values we used in our analysis). The remaining columns give the posterior means and credible intervals obtained by using the
Reversible Jump MCMC method. We summarize our results from the basic uncertain pooling method and DPmeta first, adding brief
comments about the results obtained by using the RJMCMC method at the end of this section. In general, the results are
consistent.

\begin{table}[h!t!b!p!]

\centering
\caption{\emph{Sample effect sizes, standard errors, and posterior summaries from the basic uncertain pooling
and DPmeta methods and binomial-beta model for five  COVID-19 studies \citep{he20} of the asymptomatic infection rate.}}

\setlength\tabcolsep{2pt}
\begin{tabular}{ccccccc}
  & \multicolumn{4}{c}{InvBeta Prior for $\delta^2$} & \multicolumn{2}{c}{DPmeta}  \\
 Study   &  Effect   &  PostMean    &  SE   &   $95\%$ Cred Int  &   $M=1/5$ & $ M=5$  \\
  \hline
1  & 0.310   & 0.337  &  0.128   &   (0.139, 0.644)            & 0.292&  0.348\\
2   &0.565  &  0.582  &  0.103   &     (0.354, 0.766)    & 0.655 & 0.573 \\
3  & 0.217   &0.224   & 0.045    &    (0.144, 0.322)    & 0.246 & 0.238\\
4 & 0.667  & 0.663  &  0.061   &    (0.535, 0.779)    & 0.701 & 0.662 \\
5 & 0.783  &  0.776  &  0.032    &    (0.706, 0.837)  &  0.743 & 0.769 \\
   \hline
    &   &   & &              &   &  \\
    & \multicolumn{4}{c}{Reversible Jump MCMC}  & & \\
    Study  &      &  PostMean    &     & $95\%$ Cred Int  &     &   \\
    \hline
  1  &    & 0.273  &   &    (0.145, 0.566) & &  \\
  2   &    & 0.650   &    &     (0.445, 0.780) & &  \\
  3  &     &  0.233  &      &   (0.152, 0.325) & & \\
  4 &       &   0.682  &   &    (0.547, 0.780) & & \\
  5 &      &    0.759 &   &    (0.689, 0.827) & & \\
     \hline
\end{tabular}

\end{table}



\begin{figure}[h!t!b!p!]
    \centering
\includegraphics[height=3.5in, width=4.5in]{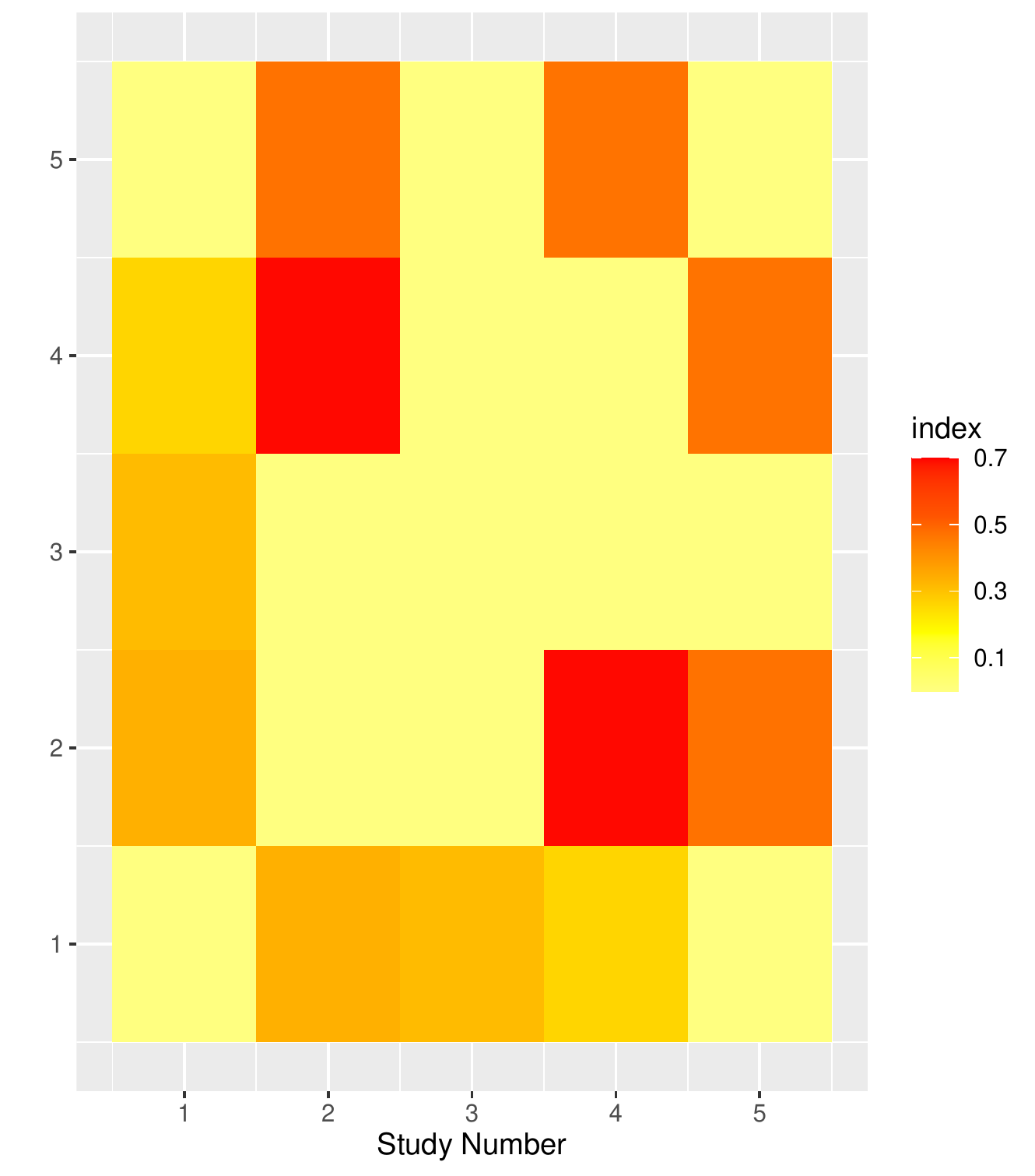}
\includegraphics[height=3.5in, width=4.5in]{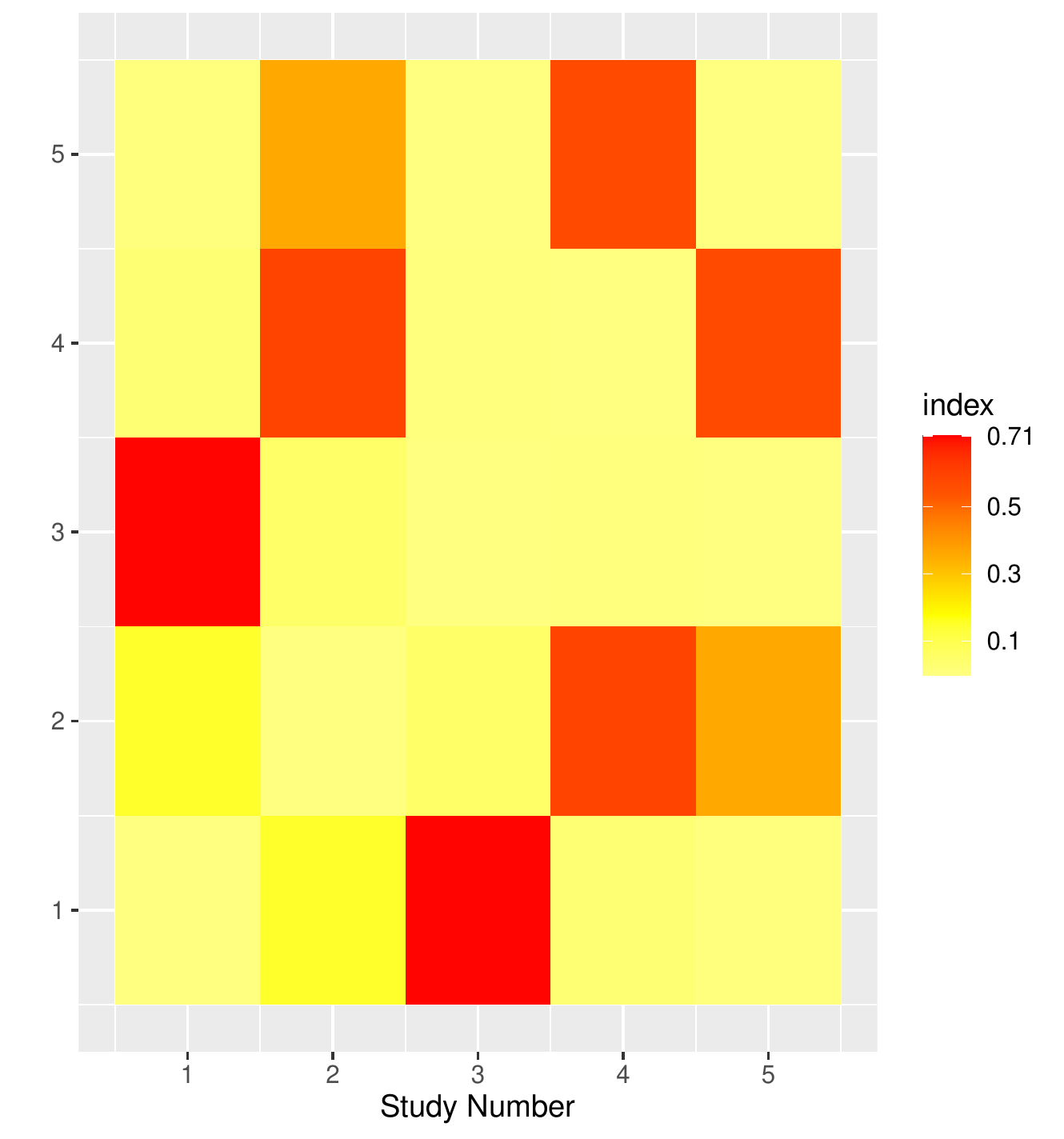}
\caption{Cell ($i, j$) gives the categorized posterior probability that studies $i$
and $j$ are in the same cluster. (Top) Basic Uncertain Pooling;  (Bottom) RJMCMC.}

\end{figure}


The posterior probability, $p(g_0 | y)$, associated with the pool-all partition, $g_0 = \{1,2,3,4,5\},$ is minuscule, i.e., $4 \times 10^{-6}.$ Now suppose that we consider the probability that there is any single
large cluster. That is, we include all partitions where $g = \{1,2,3,4,5\}$ and  $g= \{(i_1, i_2, i_3, i_4), (i_5)\}$ for $ i_j \in \{1, 2, 3, 4, 5\}$ where $i_j \neq i_k$ for $j \neq k.$ The sum
of these probabilities is $1.1 \times 10^{-4}$, a very small quantity.

A standard way \citep{gel06} to assess the likelihood that the true effect sizes come from a common source is to assume the pool-all
model, $g = g_0$, and evaluate the posterior predictive p-value using a standard   \citep[6.4 of][]{gel06},  discrepancy measure. With overall effect $\nu,$ define $y = (\overline{Y}_1,..., \overline{Y}_L)^t$ where $\overline{Y}_i$ is now the sample log odds, $\text{log} [\hat{p}_i/(1 - \hat{p}_i)],$  $\sigma^2_i/n_i$ is
$1/[n_i p_i (1 - p_i)]$ as in (\ref{transformed notation}), and $\hat{\sigma}^2_i/n_i$ is $1/(n_i \hat{p}_i (1 - \hat{p}_i))$.
The discrepancy measure,
\begin{equation}\label{Discrepancy measure}
T(y, \nu, \delta^2) = \sum_{i=1}^L \frac{(\overline{Y}_i - \nu)^2}{((\hat{\sigma}^2_i/n_i) + \delta^2)},
\end{equation}
is based on the pool-all model, as defined in (\ref{sample mean}) and (\ref{first stage prior}).
Then the posterior predictive p-value is
$$Pr\{T(y^{rep}, \nu, \delta^2)\geq T(y^{obs}, \nu, \delta^2) | y^{obs}\}$$
with $y^{rep}$ denoting a replication from the pool-all model. For these data the p-value is $3.3 \times  10^{-5}$, showing that the observed
data are not concordant with the pool-all model.
These results show that it is highly unlikely that the five true effect sizes come from a single source.

Another way to analyze these data is to construct a similarity matrix. For each pair of studies, $i$ and $j$, the similarity matrix gives the posterior probability that $i$ and $j$ are in the same cluster.
We present in Figure 1 (top) a visual representation of the similarity matrix using the basic uncertain pooling method (Section 3.1) while
Figure 1 (bottom) is the corresponding similarity plot using the RJMCMC methodology in Section 3.3.
The study numbers are given on both the $x$ and $y$ axes and there is a legend showing, for each cell, the value of the pairwise probability.
Figure 1 (top) shows  limited strong clustering, \{2,4\}, while Figure 1 (bottom) shows some additional clustering, \{1,3\},
\{2,4\} and \{4,5\}. Thus, limited aggregation seems appropriate. This may not be surprising since the  \cite{he20} review only included observations from the very early part of the COVID-19 pandemic.

From Table 1 the 95\% credible intervals are very wide, a further indication of considerable uncertainty.
 Moreover, a 95\% credible interval for the \textit{overall} true effect size, $\nu$, is $0.09 \le \nu \le 0.91$.
This wide interval is a strong indication that a single summary value such as the posterior mean of $\nu$ would not be informative.

The results from applying DPmeta provide additional insight. When $M$ is small, e.g., $M$ = 1/5, the posterior expected values show considerable clustering, i.e., \{2,4,5\} and \{1,3\}, similar to those seen in Figure 1(bottom). This clustering is not surprising since choosing a very small value of $M$ means that there is very limited sampling from the parametric centering measure $H_0$.
Using a moderate choice, $M$ = 5, the five posterior expected values are close to those obtained using the basic uncertain
pooling methodology as are the 95\% credible intervals (intervals are not shown in Table 1).

The results in Table 1 and Figure 1 suggest that one should not pool the data from these five studies. The next review that we consider,    \cite{heb20} also shows the same issue, i.e., questionable pooling of the data from all of the eleven studies. However, in this case, our analysis provides evidence of considerable clustering.  Looking at the characteristics of the eleven studies could reveal the reasons for this clustering, and the direction to take to make appropriate inferences.


The analysis presented below uses a subset of the data in \cite{heb20},
namely the asymptomatic infection rate in eleven studies of children.
The results are summarized in Table 2 and Figure 2, analogous to Table 1 and Figure 1.

\begin{table}[h!t!b!p!]
\setlength{\tabcolsep}{0pt}
\renewcommand{\arraystretch}{0.75}

\centering
\caption{\emph{Sample effect sizes,  standard errors, and posterior summaries from the basic uncertain pooling
and DPmeta methods and binomial-beta model for eleven COVID-19 children studies \citep{heb20} of the asymptomatic infection rate.}}

\setlength\tabcolsep{2pt}
\begin{tabular}{ccccccc}
  & \multicolumn{4}{c}{InvBeta Prior for $\delta^2$} & \multicolumn{2}{c}{DPmeta}  \\
Study  &  Effect   &  PostMean    &  SE   & $95\%$ Cred Int  &   $M=1/6$ & $ M= 6$  \\
  \hline
1  & 0.129 &    0.132   &       0.012&     (0.109, 0.157)   & 0.139& 0.133  \\
2   & 0.158 &  0.157     &       0.027 &    (0.114, 0.220)     & 0.142 &  0.155 \\
3  &  0.530 &  0.526 &            0.046 &    (0.436, 0.613)      & 0.507 & 0.516 \\
4  &  0.278 & 0.278   &         0.074&    (0.134, 0.489)    & 0.231 & 0.271 \\
5 &  0.129  &  0.156        &   0.060 &     (0.066, 0.308)    &0.144  & 0.159 \\
6 &    0.500  &  0.481   &        0.125 &    (0.247, 0.682)      & 0.497 & 0.455   \\
7   &  0.571 &   0.516  &       0.132 &    (0.282, 0.736)     & 0.501 &  0.482 \\
8  & 0.154 &   0.203  &        0.100  &   (0.065, 0.519)      &  0.163 & 0.192 \\
9  &   0.200 & 0.243   &       0.126 &   (0.082, 0.557)    &  0.188 & 0.221\\
10 &     0.111 & 0.210  &       0.104&    (0.037, 0.554)    & 0.173 &  0.197 \\
11 &   0.556 &  0.487    &         0.165   &      (0.191, 0.753)    & 0.488  & 0.449 \\
   \hline
 &   &   & &  &               \\
   & \multicolumn{4}{c}{Reversible Jump MCMC}  & & \\
  Study  &     &  PostMean    &   & $95\%$ Cred Int  &     &    \\
  \hline
1  &    & 0.133    &           &  (0.109, 0.159)  & &  \\
2   &    &  0.157    &           & (0.112, 0.215)  & &  \\
3  &     &  0.523     &         &    (0.430, 0.612) & & \\
4  &      &  0.272   &             &   (0.131, 0.489)  & & \\
5 &     & 0.153   &           &    (0.084, 0.257)  & & \\
6 &      & 0.490    &            &    (0.274, 0.658)  & & \\
7   &   &  0.524    &            &     (0.276, 0.726)     & &  \\
8  &   &   0.174     &       &     (0.084, 0.378)      &  &  \\
9  &     & 0.223     &            &   (0.088, 0.526)    &  & \\
10 &     &   0.171    &     &  (0.071, 0.411)    &  & \\
11 &    &  0.518     &        &   (0.266, 0.724)            &   & \\
   \hline
\end{tabular}
\end{table}



\begin{figure}[h!t!b!p!]
 \centering
\includegraphics[height=3.5in, width=4.5in]{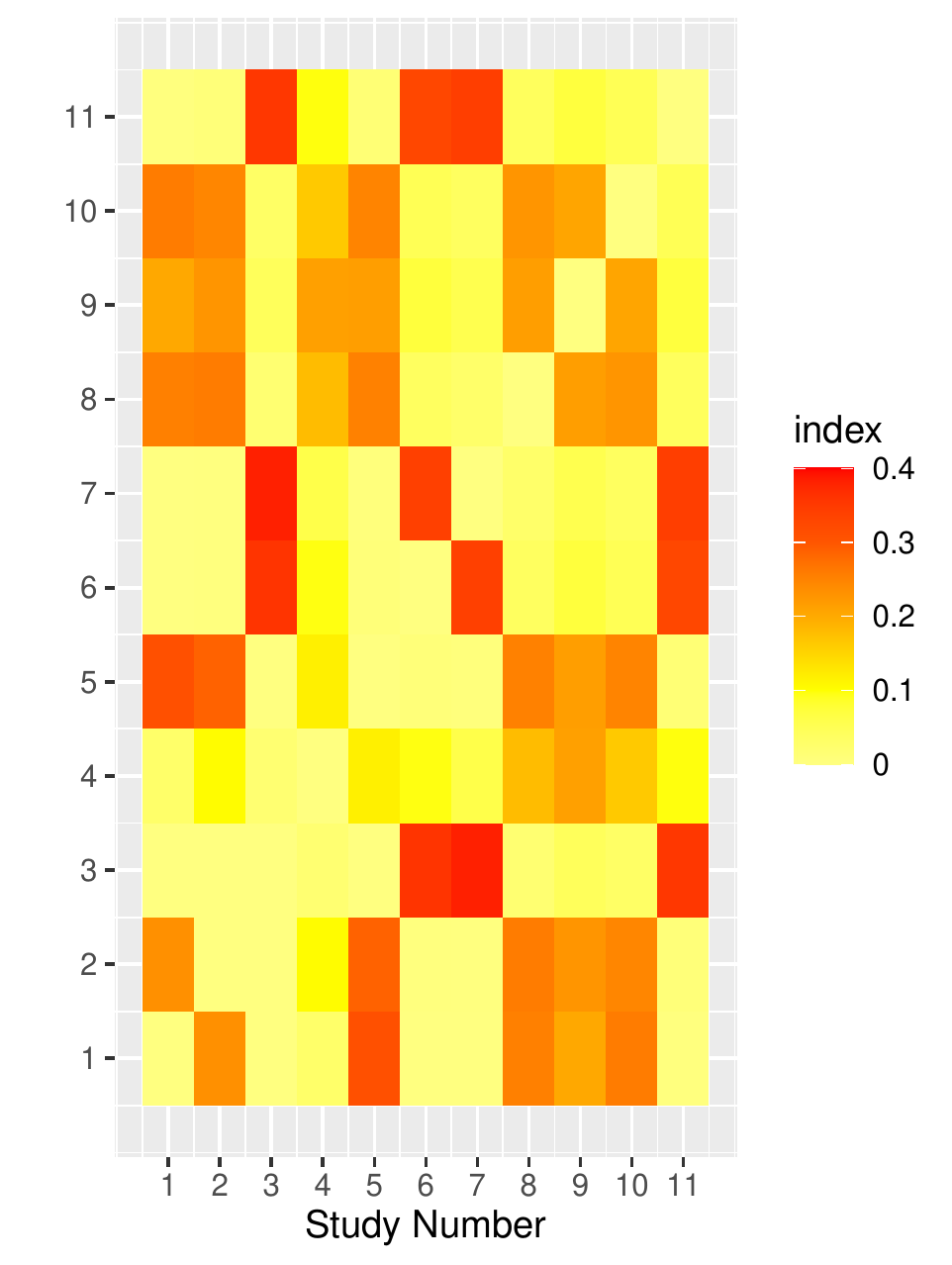}
\includegraphics[height=3.5in, width=4.5in]{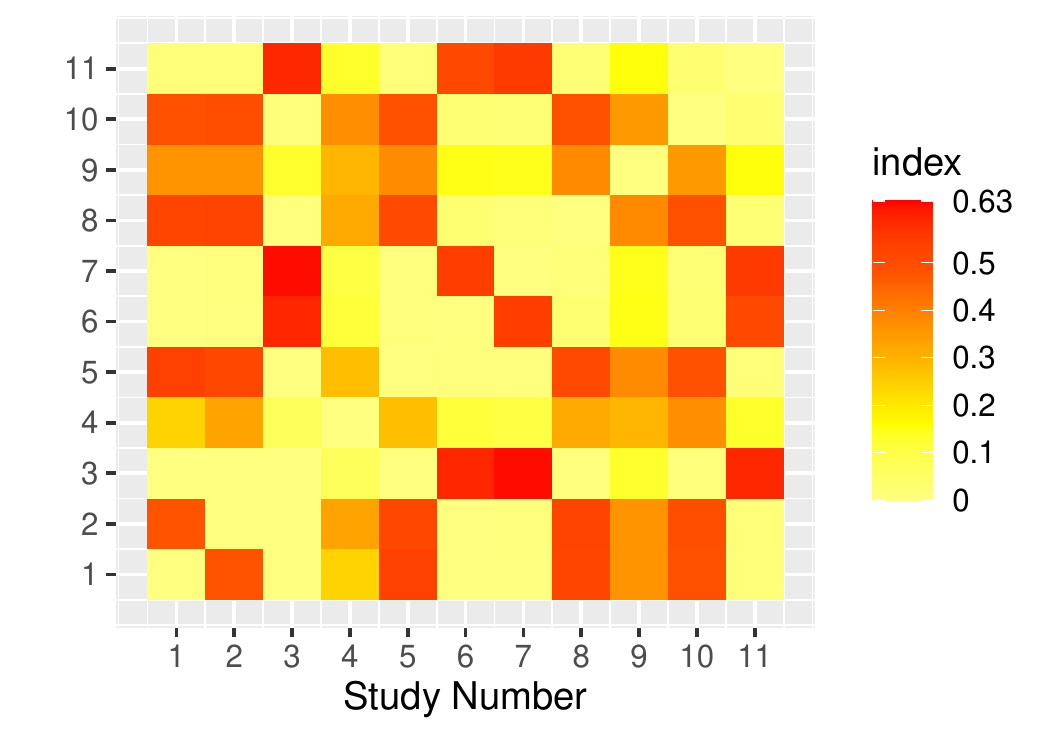}

\caption{Cell ($i, j$) gives the categorized posterior probability that studies $i$
and $j$ are in the same cluster.  (Top) Basic Uncertain Pooling;  (Bottom) RJMCMC.}
\end{figure}

From Table 2 and Figure 2 it is apparent that there are several distinct subsets. Further investigation could reveal features that separate
these subsets, leading to advances in understanding the differences in the asymptomatic rates.

The posterior probability, $p(g_0 | y)$, associated with the pool-all partition, is minuscule, i.e., $1.5 \times 10^{-11}.$
As described for the  \cite{he20}  study,
the sum of the posterior probabilities associated with partitions having only a single cluster and 0, 1, 2, 3, 4 or 5 singleton subsets is
$1.63 \times 10^{-4}$. This result suggests that it is unlikely that the eleven true effect sizes come from a single source.
Moreover, using the discrepancy measure in (\ref{Discrepancy measure}), the posterior predictive p-value is $3.3\times 10^{-5}$,
showing that the observed data are not concordant with the pool-all model. Finally, a 95\% credible interval for the \textit{overall} true effect size, $\nu$,
is $0.15 \le \nu \le 0.45$. This interval, together with the clustering, is a strong indication that a single summary value such as the
posterior mean of $\nu$ would not be informative.

From Figure 2 the most likely cluster is \{3,6,7,11\}, while the next most likely one is  \{1,2,5,8,9,10\}.
In this case, there is considerable clustering but it does not extend to all eleven studies. Without additional evidence this analysis suggests conducting separate standard meta-analyses for the two large subsets.

In Table 2 the posterior expected values for the individual studies in cluster \{3,6,7,11\} are, essentially, an average of the corresponding four sample effect sizes, but reduced in magnitude. This reduction reflects the contributions from the data from the other seven studies.

The clustering seen in Figure 2 is also evident in the posterior expected values from DPmeta with small $M$;
see Table 2 with $M = 1/6$. With $M = 6$ there is good agreement between the posterior expected values
for the two methods. For most of the eleven studies there is good agreement in the credible intervals for
the two methods with $M = 6$ for DPmeta (not shown in Table 2).

The next analysis uses the data from a subset of six of the eleven studies in  \cite{heb20}.   These six studies were chosen to
illustrate properties of the methodology when there is considerable separation.
The results are summarized in Table 3 and Figure 3.

\begin{table}[h!t!b!p!]
\renewcommand{\arraystretch}{1}

\centering
\caption{\emph{Sample effect sizes, standard errors, and posterior summaries from the basic uncertain pooling
and DPmeta methods and binomial-beta model for six COVID-19  children studies \citep{heb20} of the  asymptomatic infection rate.}}
\setlength\tabcolsep{2pt}
\begin{tabular}{ccccccc}
  & \multicolumn{4}{c}{InvBeta Prior for $\delta^2$} & \multicolumn{2}{c}{DPmeta}  \\
 Study  &  Effect   &  PostMean    &  SE   &    $95\%$ Cred Int  &   $M=1/6$ & $ M= 6$  \\
  \hline

1  &  0.129   &      0.132 &  0.012 &       (0.109, 0.157)            & 0.135 & 0.132   \\
2   & 0.158  &   0.150 &  0.027  &       (0.113, 0.211)    & 0.137  &0.152   \\
5  & 0.129  &     0.143 & 0.060   &      (0.066, 0.260)    & 0.138 &0.152  \\
6  &   0.500  &    0.521 & 0.125 &       (0.307, 0.708)  & 0.501  & 0.461  \\
7 & 0.571  &    0.548 &  0.132  &       (0.343, 0.747)    & 0.504& 0.492 \\
11 & 0.556   &   0.537 &  0.165  &      (0.265, 0.769)  &   0.495 & 0.463 \\
   \hline
   &   &   & &  &            &    \\
   & \multicolumn{4}{c}{Reversible Jump MCMC}  & & \\
  Study  &     &  PostMean    &     & $95\%$ Cred Int  &     &   \\
  \hline
1  &    &    0.132 &     &    (0.109, 0.157) & &  \\
2   &     &     0.150 &    &     (0.109, 0.200) & &  \\
5  &    &   0.142&    &       (0.082, 0.220) & & \\
6  &      &   0.528 &    &      (0.345, 0.705)  & & \\
7 &      &    0.536&    &      (0.342, 0.716) & & \\
11 &     &  0.546 &     &     (0.359, 0.741) & & \\
   \hline
\end{tabular}
\end{table}


\begin{figure}[h!t!b!p!]
 \centering

\includegraphics[height=3.5in, width=4.5in]{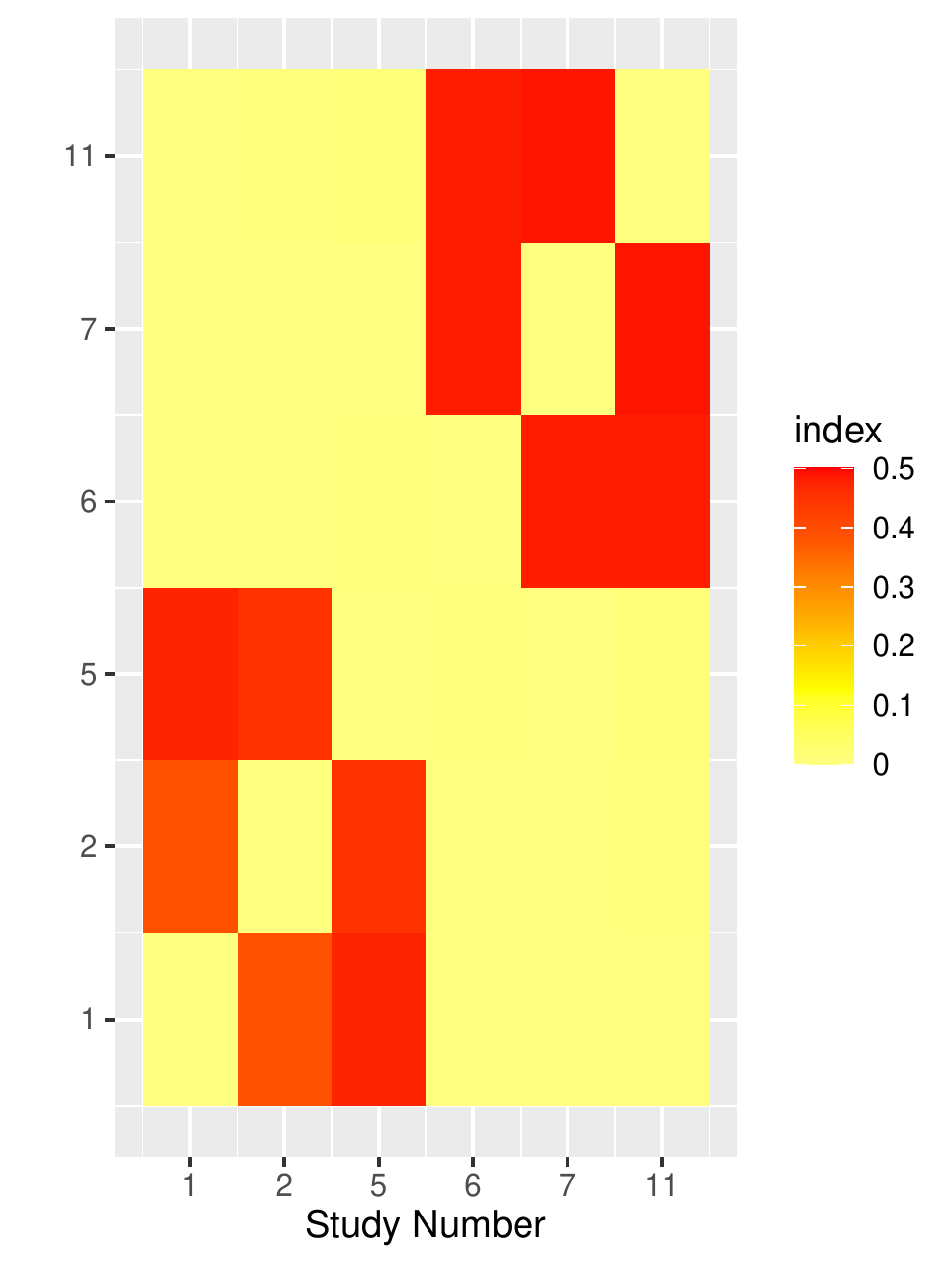}
\includegraphics[height=3.5in, width=4in]{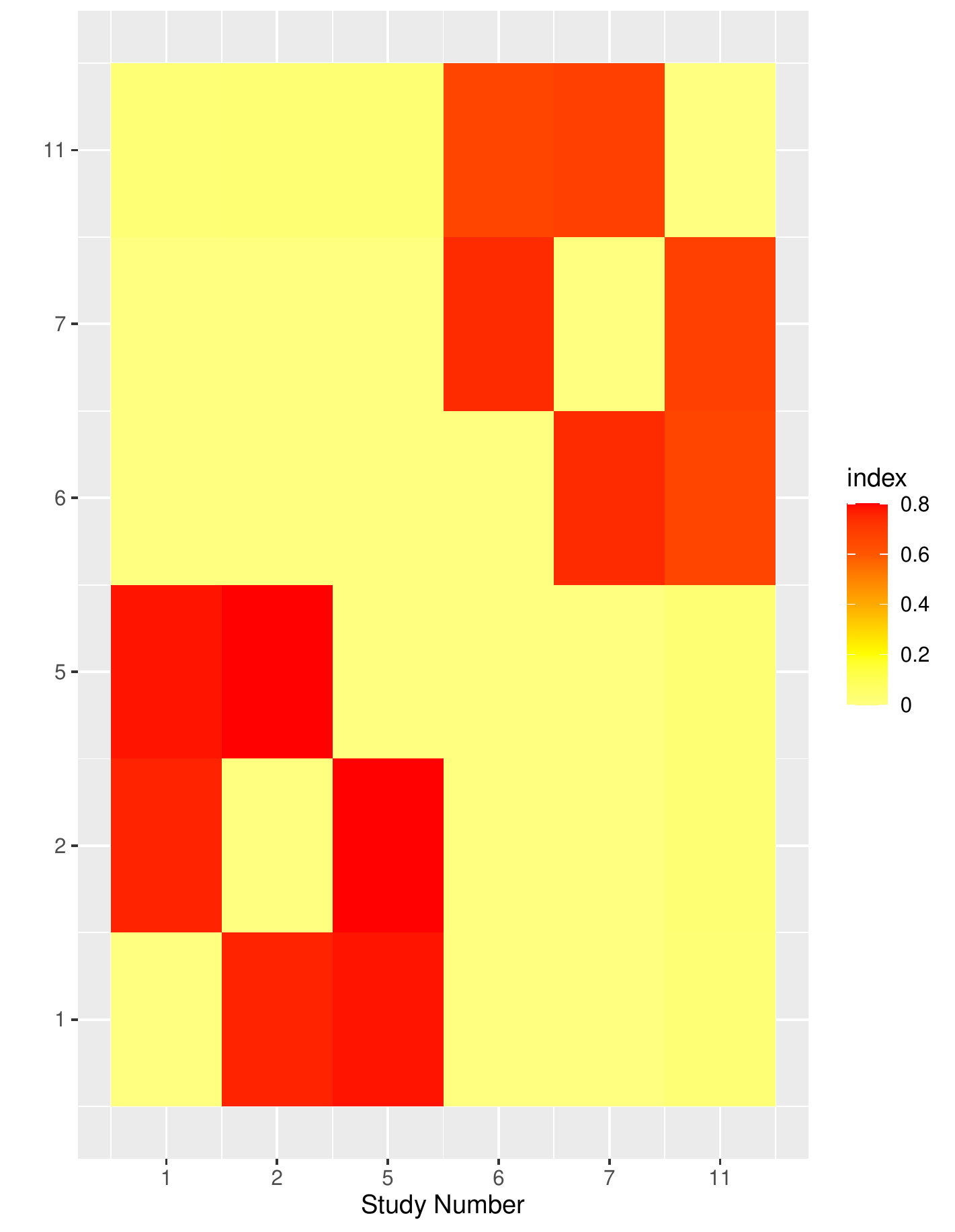}
\caption{Cell ($i, j$) gives the categorized posterior probability that studies $i$
and $j$ are in the same cluster.  (Top) Basic Uncertain Pooling;  (Bottom) RJMCMC.}
\end{figure}


From Figure 3 and Table 3 it is apparent that there are two distinct subsets, i.e., $\{1,2,5\}$ and $\{6,7,11\}$. Presumably
this separation reflects different characteristics of the two populations and/or different ways that the studies were carried
out. As expected, $p(g_0 | y)$ is minuscule, i.e., $3.1 \times 10^{-6}.$
Proceeding as described for the \cite{heb20} study, the sum of the posterior probabilities associated with partitions having only a single
large cluster (i.e., with at least four members) is $1.1 \times 10^{-3}$.
Thus, there is no evidence that the true effect sizes from these six studies
come from a single source. Moreover, using the discrepancy measure in (\ref{Discrepancy measure}), the posterior predictive p-value is $1.3\times 10^{-4}$, showing that the observed data are not concordant with the pool-all model.
Finally, a 95\% credible interval for the \textit{overall} true effect size, $\nu$,
is $0.07 \le \nu \le 0.72$. This wide interval, together with the clustering, is a strong indication that a single summary value such as the
posterior mean of $\nu$ would not be informative.
The results from DPmeta are consistent with those given above. For a  wide range of values of $M$ the
posterior expected values for \{1,2,5\} are about 0.14 while those for \{6,7,11\} are about 0.50. With $M = 6$
the posterior expected values and intervals for \{1,2,5\} are quite close to those obtained by using the basic uncertain pooling
methodology. For \{6,7,11\} almost all of the expected values and interval endpoints are within 0.06 of those from the uncertain
pooling methodology.

The final analysis uses the data from the seven study screening subset in   \cite{bge20},  identified as being of special interest.
The results are summarized in Table 4 and Figure 4. In Table 4
the sample SEs play a large role.
That is, for the studies \{1,2\} with the large SEs the posterior expected values are much smaller than the sample effect sizes. For the studies
\{5,6,7\} with the small SEs the posterior expected values and sample effect sizes are approximately equal. Thus, apart from study 4, the posterior expected values
suggest greater commonality than was apparent from the corresponding set of sample effect sizes. However, the individual posterior credible
intervals are quite different. Using the discrepancy measure in (\ref{Discrepancy measure}), the posterior predictive p-value is
$0.40$, indicating that there may be a common source for the true effects. This result is supported by Figure 4 which suggests relatively
uniform clustering (except for studies $4$ and $6$).

Since the data in each of Tables 2-4 are for a subset of a much larger set of studies they are likely to be substantially more
homogeneous than the data in the full set of studies. For example, the seven screening studies (Table 4) are a subset of seventy-nine studies with sample
proportions ranging from 0.01 to 0.92.


Results from using DPmeta are, for the most part,
consistent with these observations. For $M \in [.001, 1]$ all of the posterior expected values are approximately 0.31. For $M = 6$ the posterior expected values range from 0.25 to 0.35, similar to the posterior means in Table 4. For $M = 6$ the intervals corresponding to
\{5,6,7\}, the studies with the smallest SEs, are similar to those from the basic uncertain pooling methodology.

\begin{table}[h!t!b!p!]
 \renewcommand{\arraystretch}{1}

\centering
\caption{\emph{Sample effect sizes, standard errors, and posterior summaries from the basic uncertain pooling
and DPmeta methods and binomial-beta model for seven  COVID-19  screening studies \citep{bge20}  of the asymptomatic infection rate.}}
\setlength\tabcolsep{2pt}
\begin{tabular}{ccccccc}
  & \multicolumn{4}{c}{InvBeta Prior for $\delta^2$} & \multicolumn{2}{c}{DPmeta}  \\
 Study  &  Effect   &  PostMean    &  SE   &      $95\%$ Cred Int  &   $M=1/6$ & $ M= 6$  \\
  \hline
1  & 0.500 & 0.377   &      0.353&    (0.119, 0.838)            & 0.310  &  0.300  \\
2   &0.500 &  0.389   &    0.250 &     (0.148, 0.782)    & 0.310   & 0.307  \\
3  &  0.333&   0.332  &     0.136 &    (0.156, 0.568)    & 0.310  &  0.300 \\
4  & 0.167 &  0.226  &      0.068 &   (0.088, 0.385)  &   0.302  &  0.253  \\
5 &  0.273 &   0.288  &    0.067  &     (0.173, 0.416)    & 0.307 &  0.285  \\
6 &  0.397 &   0.382    &   0.057  &   (0.283, 0.498)  &  0.318   & 0.351 \\
7 & 0.297  &   0.300    & 0.038  &    (0.229, 0.380)  &   0.308  & 0.296 \\
\hline
    &   &   & &  &            &    \\
   & \multicolumn{4}{c}{Reversible Jump MCMC}  & & \\
  Study  &     &  PostMean    &     &     $95\%$ Cred Int  &     &   \\
  \hline
1  &  &  0.363        &     &      (0.178, 0.779) & &  \\
2   &   &   0.376       &          &     (0.192, 0.761) & &  \\
3  &     &   0.323     &         &     (0.172, 0.511) & & \\
4  &   &   0.260    &             &   (0.118, 0.378)  & & \\
5 &     &   0.297     &            &    (0.190, 0.406) & & \\
6 &     &   0.360     &         &   (0.266, 0.481) & & \\
7 &    &  0.302     &             &    (0.232, 0.377) & & \\
   \hline
\end{tabular}
\end{table}


\begin{figure}[h!t!b!p!]
     \centering
\includegraphics[height=3.5in, width=4.5in]{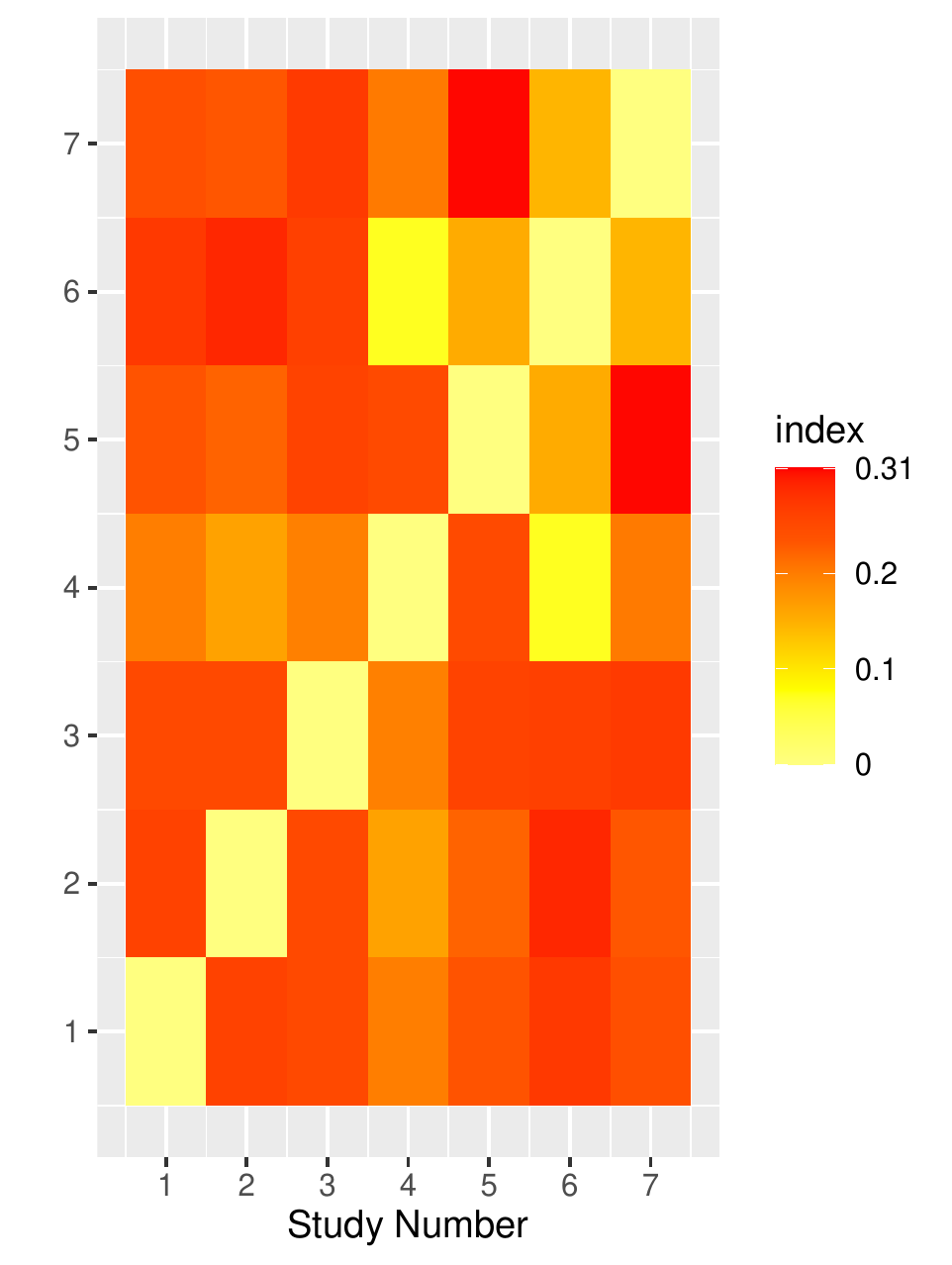}
\includegraphics[height=3.5in, width=4.5in]{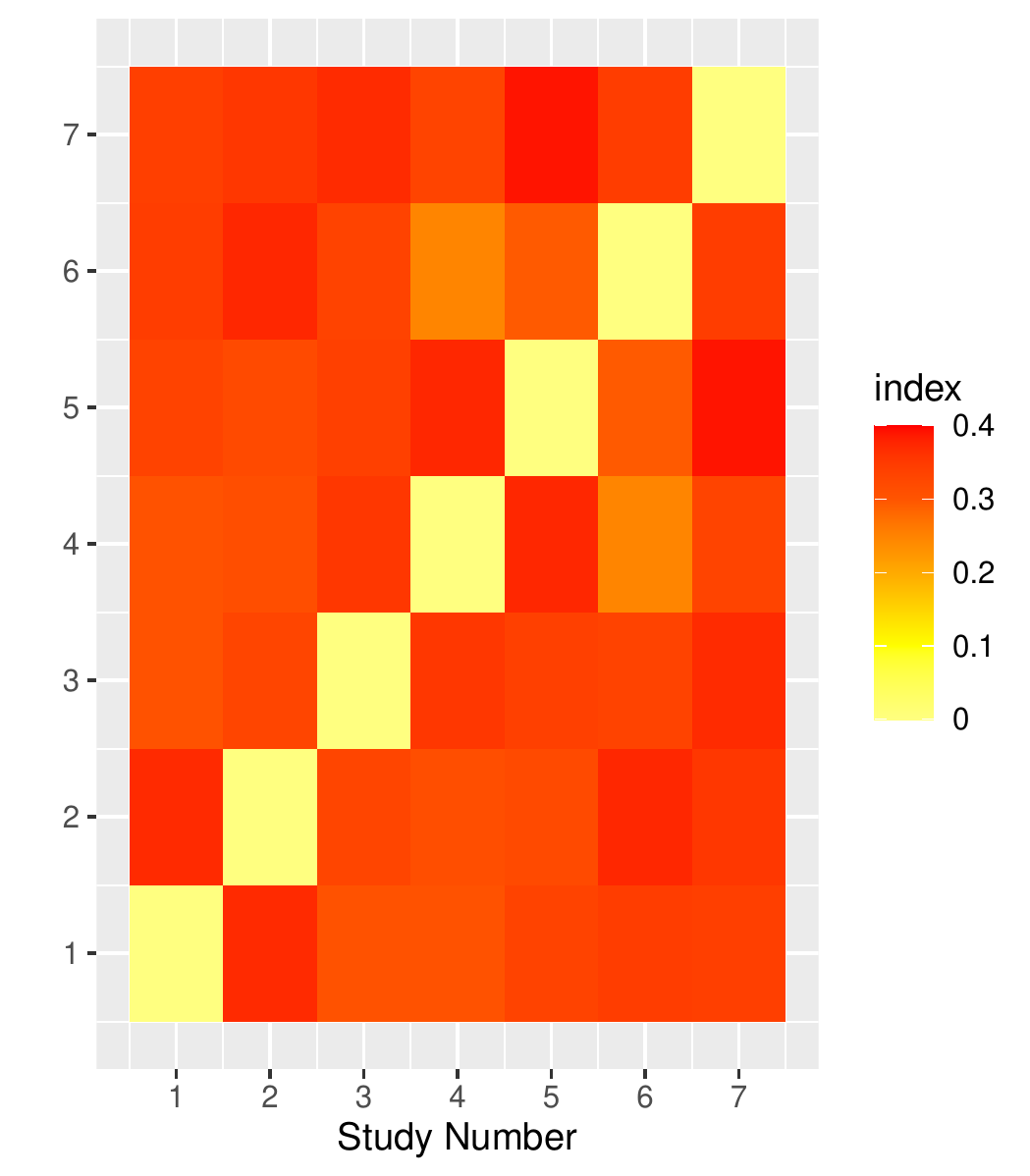}
\caption{Cell ($i, j$) gives the categorized posterior probability that studies $i$
and $j$ are in the same cluster.  (Top) Basic Uncertain Pooling;  (Bottom) RJMCMC.}
\end{figure}


With a few modifications we have implemented the RJMCMC procedure outlined in Sections 6.2 and 6.3 of  \cite{gre95}, but expand the range
for the prior for $q$ by taking log $q \sim U$(log 100, log 1000) throughout. We summarize by using the posterior means and 95\%
credible intervals (bottom of Tables 1-4), and similarity plots (bottom of Figures 1-4).

Since the sample likelihoods and prior distributions differ between the two approaches, i.e., those based on the likelihoods
and priors in Sections 3.1 and 3.3, comparisons of the results may not be especially meaningful. However, examining Tables 1-4
it is notable that the summaries (posterior means and intervals)
from the two approaches are consistent, generally close when the standard errors (SEs) are small, less so when
the SEs are very large. There are no
major differences between the comparable similarity plots corresponding to basic uncertain pooling and RJMCMC.





\section{Concluding Remarks}

The importance of good inferences for the COVID-19 asymptomatic infection rates is clear, as noted in the quotation from  \cite{bge20}  in Section 1. Conducting meta-analyses is a common, often useful, way to summarize information from a collection of studies. However, inference will be misleading if there is pooling of data from studies that are not concordant. For example,  \cite{bcb20}  note: ``A recent review by the Centre for Evidence Based medicine in Oxford found a range of estimates of asymptomatic COVID-19 cases which ranged from 5\% to 80\%. However, many of the identified studies were either poorly executed or poorly documented, making the validity of these estimates questionable.''

In this paper, we re-analyze data from three review papers, using three Bayesian methods that have a more general structure than the common meta-analytic ones. This methodology shows, in a principled manner, the extent and nature of the pooling that can be justified statistically. The more general structure should ensure greater concordance of the data with our model than with a more restricted model.

If the authors of a review have screened the studies so that the ones remaining for analysis have no markedly aberrant characteristics then an analysis showing distinct clusters should prompt a further review, and careful consideration of the inferences to make and the method to use.

In some situations there may be covariates associated with the studies that may help to explain differences in the outcomes.
To illustrate, use the basic notation and a linear regression offset. That is, replace (\ref{sample mean}) with
$$\bar{Y}_{i} \sim N(\mu_i + \mathbf{x}_i^t \beta, \sigma_i^2).$$

Then inference for $\mu$ can be made using the extension of (\ref{posterior mu})
$$f(\mu, \beta | y) = \int  f(\beta | \mu, y, g, \delta^2)f(\mu | y, g, \delta^2)f(g,\delta^2 | y)d\delta^2dg$$
where it is easily shown that
$$\beta | \mu, y, \delta^2, g \sim MVN (d, A^{-1})$$
with $z_{\mu} = (\overline{Y}_{1} - \mu_1, ... , \overline{Y}_{L} - \mu_L)^t, X$ an $L \times p$ matrix of covariates, $\beta$ a $p \times 1$ vector
of regression coefficients, $V$ an $L \times L$ diagonal matrix with $(i,i)$th element $\sigma_i^{-2}$, $A = X^tVX$, $c = z_{\mu}^tVX$ and
$d = A^{-1}c^t.$

 In the reviews we have considered,  only  \cite{heb20} gives more than one covariate for each
study, i.e., the number of confirmed cases and percent male. Using the covariates, an exploratory analysis of the residuals
showed that such an augmented analysis will not improve inferences.

The three methods can be implemented. For DPmeta there is an R package (\texttt{DPpackage}) and the code for DPmeta is included as Supplementary Material. R packages are being developed for the two uncertain
pooling methods, i.e., basic uncertain pooling and RJMCMC.


Finally, for basic uncertain pooling and DPmeta both the sample proportion and logit of the sample proportion are used in applications. While we have presented results only for the latter, our findings are similar for both choices.


 \newpage

\appendix

\section{Observed numbers of asymptomatic cases ($Y$)  and observations ($n$) \label{app1}}
\renewcommand*{\refname}{Observed number of cases ($Y_i$)  and sample sizes ($n_i$) } 

\begin{table}[h!t!b!p!]
\setlength{\tabcolsep}{18pt}
\renewcommand{\arraystretch}{1}

\centering
\caption{\emph{Observed numbers of asymptomatic cases  and observations ($Y\; / \;n$) }}
\begin{tabular*}{5.5in}{@{\extracolsep{\fill}}ccccc}
 Study  &  Table 1$^a$  &  Table 2$^b$   & Table 3$^b$    &   Table 4$^c$    \\
  \hline
1  & 4 / 13       & 94 / 728   &     94 / 728    &    1 / 2    \\
2   & 13 / 23    &  27 / 171   &    27 /  171 &   2 / 4    \\
3  &  18 / 83    &  61 / 115  &                   &      4 / 12  \\
4  &   40 / 60   &  10 /  36   &                  &       5 / 30 \\
5 &   130 / 166  &  4 / 31    &   4 / 31        &    12 / 44      \\
6 &               &  8 / 16   &       8 / 16        &          29 / 73  \\
7  &              &   8 / 14  &        8 / 14       &           41 / 138 \\
8  &              &  2 / 13   &                       &         \\
9  &              &  2 / 10   &                       &     \\
10 &             & 1 / 9      &                       &        \\
11 &               &  5 / 9    &           5 / 9     &         \\
   \hline
\multicolumn{5}{l}{\footnotesize $^a$Table 1: Effect sizes and SEs are given in Figure 4 of \cite{he20} (mislabelled as case}  \\
 \multicolumn{5}{l}{\footnotesize \hspace{0.2in}  fatality rate),  but the $n$ must be
obtained from the background papers: 1.  \cite{ni20};}  \\
\multicolumn{5}{l}{\footnotesize  \hspace{0.2in} 2.   \cite{ki20};  3.   \cite{so20};  4.  \cite{se20}; 5.   \cite{da20}.  } \\
\multicolumn{5}{l}{\footnotesize $^b$Tables 2 and 3:
$(Y,n)$ are given in Figure 3 of   \cite{heb20}. }  \\
\multicolumn{5}{l}{\footnotesize $^c$Table 4: $(Y,n)$ are given in Figure 1 of  \cite{bge20}.}
 \end{tabular*}
\end{table}



\renewcommand*{\refname}{B References} 

\section*{Acknowledgments}
The authors are grateful to the reviewers for their comments which have improved the focus
of the paper and motivated further methodological development. They are also grateful to Professor Peter Green
for his assistance in applying Reversible Jump MCMC to our data. They also appreciate
research allocation grants from XSEDE's Pittsburgh Supercomputing Center.



\subsection*{Financial disclosure}

None reported.

\subsection*{Conflict of interest}

The authors declare no potential conflict of interests.




\end{document}